\documentstyle[preprint,aps]{revtex}
\begin{document}
\draft
\title{Fractional Brownian Motion Approximation \\
Based on Fractional Integration of a White Noise }
\author{A. V. Chechkin and V. Yu. Gonchar}
\address{Institute for Theoretical Physics\\
National Science Center ``Kharkov Institute of Physics and Technology`` \\
Akademicheskaya St.1, Kharkov 310108, Ukraine }
\address{and Institute for Single Crystals\\
National Academy of Sciences of Ukraine \\
Lenin ave. 60, Kharkov 310001, Ukraine}
\date{\today }
\maketitle

\begin{abstract}
\begin{center}
We study simple approximations to fractional Gaussian noise and fractional
Brownian motion. The approximations are based on spectral properties of the
noise. They allow one to consider the noise as the result of fractional
integration/differentiation of a white Gaussian noise. We study correlation
properties of the approximation to fractional Gaussian noise and point to
the peculiarities of persistent and anti-persistent behaviors. We also
investigate self-similar properties of the approximation to fractional
Brownian motion, namely, ``$\tau ^H$ laws`` for the structure function and
the range. We conclude that the models proposed serve as a convenient tool
for the natural processes modelling and testing and improvement of the
methods aimed at analysis and interpretation of experimental data.
\end{center}
\end{abstract}


\section{Introduction.}

Fractional Brownian motion (fBm) is a random process with stationary
self-similar increments, which are normally distributed and have an infinite
span of interdependence. The general theory of random processes with
stationary self-similar increments was developed by Kolmogorov \cite{Kol-40}
as far back as 1940. Fractional Brownian motions were introduced by
Mandelbrot and van Ness as a (relatively) simple family of random functions
``that could in some way be expected to be ``typical`` of what happens in
the absence of asymptotic independence`` \cite{ManNess-68}. In this paper
the authors also have introduced fractional Gaussian noise, that is, the
stationary process, which is the derivative of the ``smoothed`` fBm. Three
classes of examples have moved these authors to study fractional motions,
namely, (i) various economic time series, (ii) $1/f$ , or flicker, noises in
electronics and solid state physics, and (iii) the tasks of hydrology, see
references in \cite{ManNess-68}. Over past 30 years the usefulness of fBm's
was confirmed repeatedly, and the number of phenomena, which can be viewed
as examples of fBm's grows. There are many papers on this subject; here we
only mention the monograph \cite{Feder-68}, which contains a lot of natural
examples, the review \cite{WestDeer-94}, which deals with the problems of
biology and physiology, and the recent monograph \cite{Mand-97}, in which
statistical properties of economical and financial time series are
discussed. We also that the description of fractional noises in terms of
probability densities requires the use of fractional calculus, which has an
ever growing interest among physicists and economists \cite{Mainardi-98}.

The ubiquity of fBm's has posed the task of developing approximations, which
allow one to simulate fBm's and study their properties in order to develop
effective methods for experimental data processing and for the purposes of
numerical modelling. Soon after the first paper on fBm \cite{ManNess-68}
Mandelbrot and Wallis have carried out simulation experiments concerning
fractional Gaussian noise \cite{MandWall-69} . They have proposed two
approximations to the noise called ``Types 1 and 2 approximations``. Later
on, Voss have proposed an empirical algorithm \cite{Voss-85} suited for
construction of not only one-dimensional fractional Brownian functions but
also fractional Brownian surfaces and volumes. However, we believe, that the
ways of constructing fractional noises and motions are not exhausted, and
various simulation models are needed, each of them may appear to be useful
when studying some particular problem.

In this paper we propose a simple model of fBm based on exploitation of
spectral properties of fractional Gaussian noise. At first we remind some
features, which are essential for our simulation and studies.

\section{Some properties of fractional Gaussian noise and fractional
Brownian motion.}

Below we use the same notations as in original papers by Mandelbrot and van
Ness \cite{ManNess-68} and by Mandelbrot and Wallis \cite{MandWall-69}. An
ordinary Brownian motion with zero mean and unit variance is denoted by $%
B(t) $, and fBm by $B_H(t)$. The latter is constructed as a moving average
of $dB(t)$, in which past increments of $B(t)$ are weighted by the kernel $%
(t-t^{\prime })^{H-1/2}$ , where $H$ is a parameter satisfying $0<H<1$. For
the ordinary Brownian motion $H=1/2$. For $H$ lying between 0.5 and 1 the
process is called persistent, whereas for $H$ lying between 0 and 0.5 it is
called anti-persistent. The introduction of fBm is motivated by the
self-similar properties of its increments, which lead to the ``$\tau ^H$
laws`` for the structure function and for the range, that is, 
\begin{equation}
S_H^{1/2}(\tau )=\left\langle \left( B_H(t+\tau )-B_H(t)\right)
^2\right\rangle ^{1/2}=\tau ^HV_H^{1/2}  \label{1}
\end{equation}
(an explicit form of $V_H$ is unessential here), and 
\begin{eqnarray}
R(t,\tau ) &=&\sup_{t\leq s\leq t+\tau }\left[ B_H(s)-B_H(t)\right] - 
\nonumber \\
&&\ \ -\inf_{t\leq s\leq t+\tau }\left[ B_H(s)-B_H(t)\right] \stackrel{d}{=}
\nonumber \\
&&\ \ \ \ \stackrel{d}{=}R(0,\tau )\stackrel{d}{=}\tau ^HR(0,1).  \label{2}
\end{eqnarray}

The properties of self-similarity of approximation to fBm will be checked
below.

It is known that the sample paths of fBm are nowhere differentiable. Such
irregular functions are best characterized by the Holder exponent or by
recently introduced local fractional derivative \cite{Kolwankar-98}. For the
purposes of our paper we note that in applications one always deals with the
``smoothed`` random process, $B_H(t,\delta )$ , which may be defined as 
\begin{equation}
B_H(t,\delta )=\frac 1\delta \int\limits_t^{t+\delta }B_H(s)ds\,\,\,,
\label{3}
\end{equation}
where $\delta \,$ is the smallest time interval, which is physically
relevant for the problem of interest. $B_H(t,\delta )$ has a derivative, 
\begin{equation}
B_H^{^{\prime }}(t,\delta )=\frac 1\delta \left[ B_H(t+\delta
)-B_H(t)\right] ,  \label{4}
\end{equation}
which is called fractional Gaussian noise. It is a stationary process with
the covariance, which may be easily determined, 
\begin{eqnarray}
C(\tau ,\delta ) &=&\left\langle B_H^{^{\prime }}(t+\tau ,\delta
)B_H^{^{\prime }}(t,\delta )\right\rangle  \nonumber \\
&=&\frac{V_H}{2\delta ^{2H}}\left\{ (\tau +\delta )^{2H}-2\tau ^{2H}+(\tau
-\delta )^{2H}\right\}  \label{5}
\end{eqnarray}
If $\tau \ll \delta $ , it follows from Eq.(5) that 
\begin{equation}
C(\tau ,\delta )\propto (2H-1)\tau ^{2H-2},  \label{6}
\end{equation}
and thus, the covariance has the same sign as $H-1/2$ and has a very
different behavior in a persistent case and in anti-persistent one. The
analysis of Eqs.(5), (6) shows that for the former case $C(\tau ,\delta )$
is positive for all $\tau $, and 
\[
\int\limits_0^\infty C_H(s,\delta )ds=\infty , 
\]
whereas for the latter $C(\tau ,\delta )$ changes sign once from positive to
negative at a value of $\tau $ proportional to $\delta $, and one has 
\[
\int\limits_0^\infty C_H(s,\delta )ds=0. 
\]
The next property, which serves as the basic one at the initial stage of our
simulation, is a power-law behavior of the spectrum of fractional Gaussian
noise, 
\begin{equation}
G(f;\delta )=4\int\limits_0^\infty C_H(s,\delta )\cos (2\pi fs)ds\propto
f^{1-2H}  \label{7}
\end{equation}
at $f\delta <<1$.

The spectrum (7) is also quite different for persistent and anti-persistent
cases. For the former the main part of the noise energy is concentrated in a
low -frequency region, whereas for the latter the spectral intensity grows
towards high frequencies.

In numerical simulation one deals with a discrete-time fractional noise,
when $t$ is a positive integer and the noise is defined by the sequence of
increments $B_H(t+1)-B_H(t)$ . The correlation function is defined by Eq.(5)
with $\delta =1$. The Fourier transform of the discrete-time noise is
defined between plus and minus the Nyquist critical frequency $f_c=0.5.$

The power-law dependence of the spectrum at small $f$ allows one to simulate
fractional Gaussian noise by fractional integration/differentiation of a
white noise, with a subsequent construction of fBm.

\section{The models.}

We remind the relation between the Fourier transforms of the function $X(t)$
and of its fractional integral/derivative $X_\nu (t)$ of the $\nu $ order: 
\begin{equation}
\stackrel{\wedge }{X}_\nu (f)=\frac{\stackrel{\wedge }{X}_\nu (f)}{(-i2\pi
f)^\nu },  \label{8}
\end{equation}
where $\nu $ is positive in case of fractional integration and negative in
case of fractional differentiation (to be more accurate, we say about
left-side Liouville fractional integral/derivative on the infinite axis \cite
{Samko-87}). If $X(t)$ is a white Gaussian noise, then the spectral density
of $X_\nu (t)$ is, at least, formally, 
\begin{equation}
G_\nu \left( f\right) \propto \frac 1{f^{2\nu }},  \label{9}
\end{equation}
which behaves as the spectrum of fractional Gaussian noise at small $f$, see
Eq.(7), if we set $\nu =H-1/2,\quad -1/2<\nu <1/2$ . Thus, we use the
following way for simulation:

(i) taking white Gaussian noise $X(t)$, $t$ is an integer, we multiply its
Fourier transform by $f^{-\nu }$, $-1/2<\nu <1/2.$

(ii) making an inverse Fourier transform, we get $X_\nu (t)$, which is
supposed to approximate fractional Gaussian noise with the index $H=+1/2$.

(iii) The process $X_\nu ^{*}(t)$ , which is supposed to approximate fBm, is
defined by 
\begin{equation}
X_\nu ^{*}(t)=\sum\limits_{\tau =1}^tX_\nu (t).  \label{10}
\end{equation}

\section{Numerical results.}

The results of numerical simulation and analysis are shown in Figs.1-5.

At the left of Fig.1 we show typical examples of the samples of the process $%
X_\nu (t)$ for (a) $\nu =0.4$, (b) $\nu =0$(white Gaussian noise), and (c) $%
\nu =-0.4$.The stationarity of the samples was checked by the methods of
series and inversions \cite{Bendat-86}. The figures demonstrate clearly the
prevalence of low-frequency components for the case (a), and the prevalence
of high-frequency components for the case (c). At the right of Fig.1 the
solid lines depict normalized theoretical correlation functions, see Eq.(5),
with $\delta =1$, for (a) $H=0.9$, (b) $H=0.5$, and (c) $H=0.1$, whereas the
normalized correlation functions 
\begin{equation}
C(\tau )=\frac{\left\langle X_\nu (t)X_\nu (t+\tau )\right\rangle }{%
\left\langle (X_\nu (t))^2\right\rangle },  \label{11}
\end{equation}
are shown by black points for the same values of as at the left of the
figure. The results demonstrate good agreement between correlation function
of simulated process with the parameter $\nu $ and theoretical correlation
function for the fractional noise with the corresponding index $H=\nu +0.5$.
We also see clear difference between correlation functions for the
persistent and anti-persistent cases. This difference is the manifestation
of the property, which at a qualitative level can be formulated as follows:
in the persistent random process the available tendency is supported,
whereas in the anti-persistent process the opposite tendency prevails \cite
{ManNess-68}\cite{Feder-68}. We also note that the correlation functions
indicated by black points are estimated from the samples shown at the left
with the use of time averaging. The complementary averaging over a set of
realizations diminishes scattering of the points.

It is known, that the wavelet transform is well-suited for analyzing
structure of non-stationary processes \cite{Gross-84}. In particular, it
allows one to study the behavior of processes at different scales and hence
it is sometimes called as a mathematical microscope. The wavelet transform $%
T(a,t)$ of the process $X_\nu ^{*}(t)$ is written as 
\[
T(a,t)=\frac 1{\sqrt{a}}\int\limits_{-\infty }^\infty W\left( \frac{%
t-t^{\prime }}a\right) X_\nu ^{*}(t^{\prime })dt^{\prime }, 
\]
where $W(t)$ is called analyzing wavelet. In Fig.2 we demonstrate the
wavelet transforms $T(a,t)$ of the samples of $X_\nu ^{*}(t)$ for (a) $\nu
=0.4$, and (b) $\nu =-0.4$. We use the ``Mexican hat`` as analyzing wavelet, 
\[
W(t)=(1-t^2)\exp (-t^2/2). 
\]
The low part of the wavelet corresponds to small $a$ values (that is, to
high frequencies), whereas the upper part corresponds to large $a$ values
(low frequencies).Below each of the wavelet the sample path is shown, which
is subjected to wavelet transform. It is seen that the wavelet for
anti-persistent case has a clearly visualized structure in a high-frequency
region. On the contrary, the wavelet for persistent case demonstrate almost
homogeneous background in a high-frequency region. The lines of local maxima
demonstrate hierarchial structure, which reflects self-similarity of fBm.

In this paper we do not intend to use wavelets for extracting quantitative
characteristics, so, Fig.2 has an illustrative purpose only. However, we
note that the wavelet transform being applied to some known functions allows
one to establish their multifractality \cite{Arneodo-95}.

We study numerically $\tau $-dependence of the structure function of the
generated process $X_\nu ^{*}(t)$ , 
\begin{equation}
S(\tau )=\left\langle \left( X_\nu ^{*}(t+\tau )-X_\nu ^{*}(t)\right)
^2\right\rangle \propto \tau ^{2s}.  \label{12}
\end{equation}
Since for fBm the ``$\tau ^H$ law`` is fulfilled, see Eq.(1), we expect that 
$s$ is close to $\nu +0.5$. In Fig.3 $s$ depending on $\nu $ is depicted by
points for the persistent processes and by crosses for the anti-persistent
ones. The expected relation is indicated by dotted line. We see that the
numerical relation is well fitted by the expected line, however, some
discrepancies appear for the strongly persistent and strongly
anti-persistent processes as well. At the insets the structure function
versus time delay $\tau $ is indicated by black points in log-log scale. The
slopes of fitted lines give the values of $s$ for $\nu =-0.4$ (top inset)
and $\nu =0.4$ (bottom inset), respectively.

We now turn to the study of the range of $X_\nu ^{*}(t)$ . In the empirical
rescaled range analysis , that is, at experimental data processing or in
numerical simulation the range of the random process is divided by the
standard deviation of its increments after subtraction of a linear trend,
see Ref. \cite{Hunt-65}. This procedure, in particular, smooths the
variations of the range on different segments of time series. As the result
of the empirical rescaled range analysis of experimental data one gets the
Hurst exponent of the process. For fBm the Hurst exponent thus obtained must
coincide with the parameter $H$, see Eq.(2). Fig.4 demonstrates the
application of the rescaled range analysis to the sample paths of $X_\nu
^{*}(t)$ . The index $\nu $ is 0.4, thus the value of the Hurst exponent,
which we expect to get according our way of simulating fBm is 0.9. In Fig.4a
the fluctuations of the range $R$ (thin curve) and those of the standard
deviation $\sigma $ of the increments (thick curve) are shown for the case
when the total length of the sample is divided into 64 segments, each of =16
lengthwise. Below the variations of the ratio $R/\sigma $ are depicted. It
is shown that fluctuations of the ratio are smaller than those of the range.
This circumstance justifies the use of the ratio in the empirical analysis.
In Fig.4b the rescaled range versus time interval $\tau $ is depicted by
black points in log-log scale. The slope of the fitted line gives the Hurst
exponent 0.83.

In Fig.5 the values of the Hurst exponent $H_\nu $ obtained by applying the
rescaled range analysis to the sample paths of $X_\nu ^{*}(t)$ are shown by
black points for the persistent processes and by crosses for the
anti-persistent ones. The dots indicate the line $H_\nu =\nu +0.5$, that is,
the relation, which we expect to get according our way of simulating fBm. It
is shown that (i) fairly good agreement is for intermediate $\nu $, (ii)
deviations between the results for $X_\nu ^{*}(t)$ and the theory for fBm
grow to the ends of the interval of $\nu $, and (iii) these deviations are
larger for the anti-persistent case than for the persistent one.

\section{Discussion.}

Figures 3 and 5 allow one to conclude that in the persistent case there is a
better agreement with the expected lines than in the anti-persistent case
(with the length of the sample and discretization step being equal for both
cases). In other words, in our model the low frequency errors are less
important than the high-frequency ones. This conclusion is analogous to that
concerning ``Type 2 approximation`` constructed by Mandelbrot and Wallis 
\cite{MandWall-69}. This is not surprising since the way of its constructing
in real space is similar to our way of constructing approximation in Fourier
space. Moreover, we compare the results of our numerical simulations with
those performed with the ``Type 2 approximation`` and the Voss algorithm 
\cite{Voss-85} and found unessential differences between the results for the
exponent of the structure function, see Fig.3, and for the Hurst index, see
Fig.5, for the three methods. We may conclude that our method is suited for
performing simulation of fBm as well as the ``Type 2 approximation`` and the
Voss algorithm are. In all three models the persistent case is simulated
with higher accuracy than the anti-persistent one. Fortunately, the
persistent case is prevalent in nature \cite{Feder-68}, therefore, the
method proposed in our paper can be used in various applications. At last,
we believe that our method will be useful for direct constructing
approximations to non-Gaussian fractional noises, in particular, fractional
Levy noises. This topic will be the subject of forthcoming paper.

\acknowledgments

This work was supported in finance by National Academy of Science of
Ukraine, the Project ``Chaos-2`` and by INTAS, the Project 93-1194. The
information support within the Project INTAS LA-96-09 is also acknowledged.

\newpage

\begin{figure}[tbp]
\caption{At the left: typical samples of the simulated noise $X_n(t)$ for
(a) $\nu =0.4$; (b) $\nu =0$ (white Gaussian noise), and (c) $\nu =-0.4$. At
the right: normalized correlation functions of the simulated noises with the
same $\nu $ (black points). Solid lines indicate normalized theoretical
correlation functions, see Eq.(5) with $\delta =1$, for (a) $H=0.9$, (b) $%
H=0.5$, and (c) $H=0.1$.}
\label{fig1}
\end{figure}

\begin{figure}[tbp]
\caption{Wavelet transforms of the simulated process $X_\nu ^{*}\left(
t\right) \,\,$for (a) $\nu =0.4$, and (b) $\nu =-0.4$. Below each of the
wavelets the sample paths subjected to wavelet transform are shown.}
\label{fig2}
\end{figure}

\begin{figure}[tbp]
\caption{The exponent $s$ in the structure function of the simulated
process, see Eq.(11), versus the index $\nu $. Black points indicate
persistent case, whereas the crosses indicate the anti-persistent one. The
line $s=\nu +1/2$ is depicted by dots. At the insets: the square root of the
structure function versus time in log-log scale (black squares) for $\nu
=-0.4$ (top inset) and $\nu =0.4$ (bottom inset). The slopes of fitted lines
give the values of the exponent $s$, see the main figure.}
\label{fig3}
\end{figure}

\begin{figure}[tbp]
\caption{(a) Variations of the range of $X_\nu ^{*}\left( t\right) \,\,$
(thin curve), of the standard deviation of $X_\nu (t)$ (thick curve), and of
their ratio (below) at the different time segments. (b) Rescaled range
versus time interval in log-log scale (black points). The slope of the
fitted line gives the Hurst index 0.83.}
\label{fig4}
\end{figure}

\begin{figure}[tbp]
\caption{Plots of the Hurst exponent $H_\nu $ versus $\nu $ for the
persistent case (black squares) and anti-persistent one (crosses). The line $%
H_\nu =\nu +1/2$ is indicated by dots.}
\label{fig5}
\end{figure}

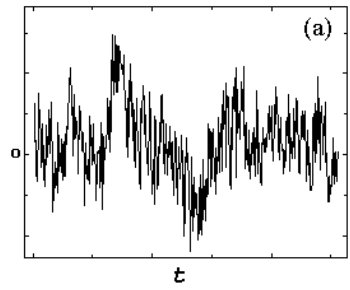
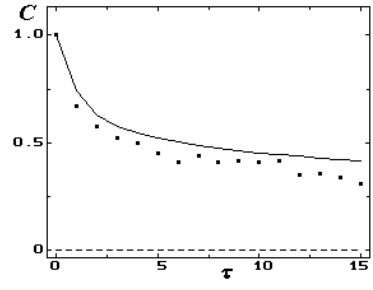
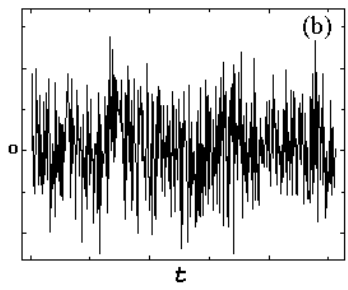
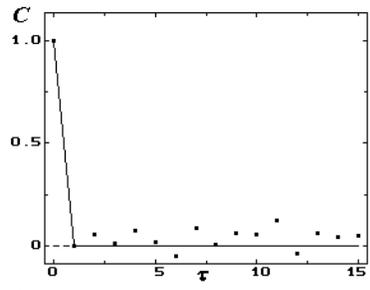
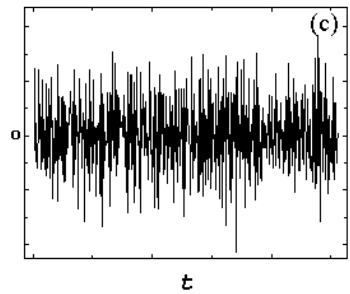
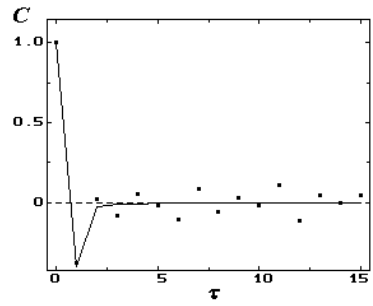
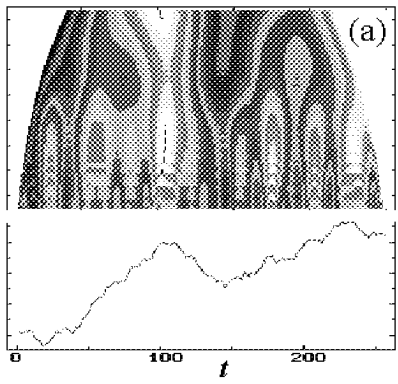
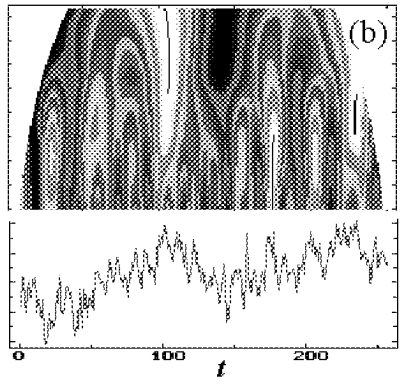
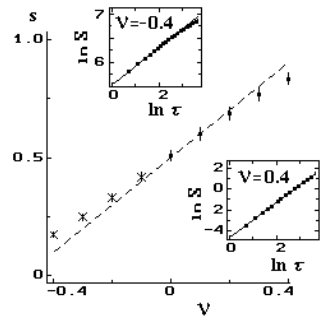
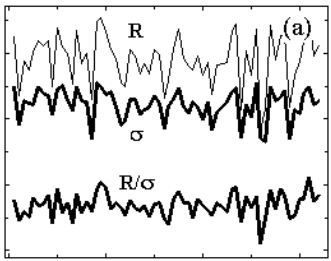
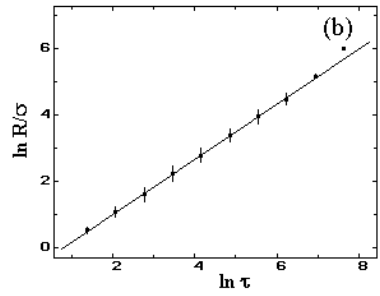
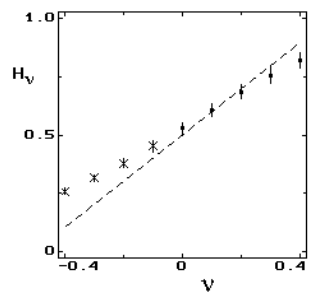
\end{document}